\documentclass[fleqn,usenatbib]{mnras}

\usepackage{newtxtext,newtxmath}

\usepackage[T1]{fontenc}
\usepackage{soul} 

\DeclareRobustCommand{\VAN}[3]{#2}
\let\VANthebibliography\thebibliography
\def\thebibliography{\DeclareRobustCommand{\VAN}[3]{##3}\VANthebibliography}

\defcitealias{BailerJones2018}{BJ18}
\defcitealias{DavenportCovey2018}{DC18}
\defcitealias{McQuillan2014}{MQ14}
\defcitealias{Amard2019}{STAREVOL}
\defcitealias{GaiaKepler}{Gaia-\textit{Kepler}}

\usepackage{graphicx}	
\usepackage{amsmath}	
\usepackage{amssymb}	
\usepackage{url}

\title[Metallicity-Dependant Spin Evolution]{Evidence for Metallicity-Dependant Spin Evolution in the \textit{Kepler} field}

\author[Amard et al.]{
Louis Amard,$^{1}$\thanks{E-mail: l.amard / j.torres-roquette / s.matt @exeter.ac.uk}
Julia Roquette,$^{1 \star}$
Sean P. Matt$^{1 \star}$
\\
$^{1}$University of Exeter, Physics and Astrophysics dept, Exeter, EX44QL, UK\\
}

\date{Accepted XXX. Received YYY; in original form ZZZ}

\pubyear{2020}

\begin{document}
\label{firstpage}
\pagerange{\pageref{firstpage}--\pageref{lastpage}}
\maketitle

\begin{abstract}
A curious rotation period distribution in the Color-Magnitude-Period Diagram (CMPD) of the \textit{Kepler} field was recently revealed, thanks to data from Gaia and \textit{Kepler} spacecraft.  It was found that redder and brighter stars are spinning slower than the rest of the main sequence.  On the theoretical side, it was demonstrated that metallicity should affect the rotational evolution of stars as well as their evolution in the Hertzprung-Rüssel or Color-Magnitude diagram. 
In this work we combine this dataset with medium and high resolution spectroscopic metallicities and carefully select main sequence single stars in a given mass range. 
We show that the structure seen in the CMPD also corresponds to a broad correlation between metallicity and rotation, such that stars with higher metallicity rotate on average more slowly than those with low metallicity.
We compare this sample to theoretical rotational evolution models that include a range of different metallicities. They predict a correlation between rotation rate and metallicity that is in the same direction and of about the same magnitude as that observed.
Therefore metallicity appears to be a key parameter to explain the observed rotation period distributions.
We also discuss a few different ways in which metallicity can affect the observed distribution of rotation period, due to observational biases and age distributions, as well as the effect on stellar wind torques.

\end{abstract}




\section{Introduction}

A star can be described from their fundamental properties, mass, chemical composition, age and angular momentum content or rotation period.
However, except for the Sun, we never know all these parameters independently and we have to exploit the few we have at hand. 
In particular, the age of stars is especially challenging to obtain and mostly rely on the modelling based on other properties. Our knowledge of the stellar mass and chemical composition can help to retrieve individual stellar ages with typical isochrone fitting or with more advanced techniques like asteroseismology  \citep[for a review on age determination techniques see][and references therein]{Soderblom2010}. For low-mass stars ($T_\textrm{eff}\lesssim 6250$K), rotation period can also be a key component to determine an accurate age via a technique called gyrochronology \citep{Barnes2003}. 
It is made possible by the systematic and well constrained spin-down of low-mass stars along their main sequence life \citep{Skumanich1972}, despite the initial large range of rotation periods with which stars are born. Stellar winds couple with the large scale magnetic field and exert a torque on the stellar surface mostly to spin-down the star as the material is going away \citep{Parker1958,Schatzman1959,WeberDavis1967,Mestel1984,Kawaler1988,Krishnamurthi1997,Matt2012,VSP13,GalletBouvier2013}.
The presence of a deep convective envelope in lower mass stars allows the generation of a large scale magnetic field thanks to a stellar dynamo and lead to a continuous loss of angular momentum by magnetised stellar winds during the main sequence.
This magnetic braking mechanism is thus dependent on the ability of the star at generating a large scale magnetic field. This ability is usually described by the ratio of the inertial force and the Coriolis force, which is here given by the stellar Rossby number defined as the ratio between the rotation period and the convective turnover timescale \citep[\textit{e.g.}][for a review]{BrunBrowning2017}. The smaller the Rossby number, the stronger the magnetic field, up to a certain saturation level. The size of the convective envelope and more generally the structure of the star are strongly affected by the chemical composition of the star \citep[See for example][]{Kippenhahn2012}. It was recently shown how a small variation in the amount of elements heavier than helium in a star can modify the rotational evolution of a star, mostly because of the induced change in opacity and its effect on the stellar structure \citep{Claytor2020,AmardMatt2020}. 

\citet{VanSaders2016} and Hall et al. (subm) did a complete study with \textit{Kepler} and K2 asteroseismic targets of which they have well-constrained ages, masses, metallicity and rotation period and present some constraints on the stellar wind torque. \citet{LorenzoOliveira2019} tested the rotation evolution scenarios and determined masses and ages based on empirical scalings.
However, these samples are at most of a hundred stars and until now, no large sets of data were complete enough to provide the chemical composition, the rotation period and some constrains on the stellar mass (without even mentioning the age). \citet{VanSaders2019} for example compared the \textit{Kepler} field rotation periods distribution to their models in the rotation period-effective temperature plane without knowing the chemical composition of the stars. The latter was accounted for assuming the metallicity distribution was following the one from a galactic model with all the uncertainties it comes with. It allowed them to draw some conclusions on the existing biases when determining stellar populations ages. For example, they show that the stars with a detected rotation period observed by \textit{Kepler} are strongly biased after a given age limit, which depends on their effective temperature, higher temperatures being associated with a lower age limit.
\cite{Claytor2020} presented the first consistent determination of stellar ages of \textit{Kepler} field stars using gyrochronology, and accounting for the stellar chemical composition of each star to compute their rotational evolution, they compiled a sample of about 500 stars.

The \textit{Kepler} mission provided light curves for several hundred of thousands of stars, of which \cite{McQuillan2014} (hereafter \citetalias{McQuillan2014}) was able to extract the rotation period for about 34000 low-mass stars. 
Then, the Gaia mission provided magnitudes, colors and accurate distances for the largest number of stars ever observed and allowed to display them in color-magnitude diagrams  \citep{Babusiaux2018}.
\citet[][hereafter \citetalias{DavenportCovey2018}]{DavenportCovey2018}, following \citet{Davenport2017}, showed for the first time with respectively Gaia DR2 and DR1 data, what the \textit{Kepler} field stars with known rotation periods are looking like in a colour-magnitude diagram. They reveal in particular a difference in rotation rates between stars on parallel tracks on the diagram. They could not find any physical explanation and only demonstrate that the broadening caused by an age spread is not sufficient to explain such feature.
Indeed, although an age range allows to cover all rotation periods, it does not broaden the main sequence uniformly for all masses (see Appendix).
Metallicity is known to have a comparable effect to stellar mass or age and globally defines the path along which stars will evolve in the CMD \citep{HS1955,Kippenhahn2012}.
However, \citetalias{DavenportCovey2018} originally dismissed metallicity as an explanation for this feature.
Their assumption was that metal-enriched stars, expected to be younger on a galactic scale, should thus be faster rotators, which is the opposite trend to that observed.
Galactic archaeology and models of the Galaxy suggest however that in the last 8Gyr, the median metallicity as well as the spread have not changed much \citep{Haywood2013,Haywood2019}. Moreover, there was not enough reliable metallicity data available at that time to be used with such a large sample.
In the last few years, several large scale medium to high resolution spectroscopic surveys have emerged. Among these, the most prominent is LAMOST \citep{Luo2015} which has observed a few millions targets and provided reliable metallicity and effective temperature for a good third of the above mentioned Gaia-\textit{Kepler} sample. 

In this work, we demonstrate the importance of both the stellar mass and the chemical composition of a stellar population to comprehend the underlying physics while low-mass stars are spinning down on the main sequence.
We use the \textit{Kepler} field as an example to show the effect of a spread in metallicity on the rotation period distribution of a stellar population. 
We carefully select the best data to date for the study, remove most known biases and describe the obtained sample in \S~\ref{Sec:Obs}. In \S~\ref{sec:Zinfluence} we study in detail the metallicity distribution and its effect on other stellar parameters. We compare our observation sample to existing models of rotating stars population in \S~\ref{sec:comp_synth}. Finally, we discuss our results and conclude in sections \ref{sec:discussion} and \ref{sec:conclusion}.

\section{The observational Sample and Analysis}
\label{Sec:Obs}

  \subsection{Full sample with photometry, distances, periods, and metallicities} \label{sec:fullsample}

    \subsubsection{Rotational periods and Gaia photometry and distances}\label{sec:periodsGaia}

We used the \citetalias{GaiaKepler} catalogue as our starting point, which includes entries for 201,312 sources in the \textit{Kepler} field and was produced by a cross-matching between Data Release 25 \textit{Kepler} Catalog and Gaia DR2 source catalogue, using a 1 arcsecond radius for matching, and includes the improved distance prescription from \cite{BailerJones2018} (hereafter \citetalias{BailerJones2018}). 
We then excluded duplicate sources (removing those with \verb|duplicated_source=True|) and selected high quality Gaia-DR2 data by requiring a parallax error $<0.1$ mas (\verb|parallax_error<0.1|) and $\frac{\sigma_ m}{m}<0.01$ for every photometric band (\verb|phot_X_mean_flux_error/phot_X_mean_flux<0.01|, for \verb|X=BP, RP, G|).  
We converted Gaia apparent magnitudes to absolute magnitudes ($M_G=G-5\log_{10}(\frac{d_\mathrm{BJ18}}{10})$) by using the distances from \citetalias{BailerJones2018} as a substitute to a simple inversion of Gaia parallaxes to obtain distances. 
Following the suggested use of their source catalogue, we only used sources which did not have a bimodal distance solution (\verb|modality_flag == 1|) and that had a well-constrained distance (\verb|result_flag == 1|) in Gaia DR2.  Finally, we merged this \citetalias{GaiaKepler} sample with the rotation period measurements from \citetalias{McQuillan2014}, using the target identification number in the \textit{Kepler} Input Catalogue (hereafter KIC).  This process results in a sample of 28\;508 stars with good quality Gaia DR2 data and measured \textit{Kepler} rotation periods.  
More recent and extended sets of rotation periods data are available \citep[\textit{e.g.}][]{Santos2019}, however, we have opted to use the original one since this provides the most homogeneous parameters.
We checked that completing our sample with theirs does not alter the results of this paper.

    \subsubsection{Metallicity data} \label{sec:metallicity}

We searched the literature for measurements of [Fe/H] for stars in the \textit{Kepler} Field that were based on mid- or high-resolution spectroscopy, ultimately selecting the catalogues from the Revised \textit{Kepler} Input Catalogue \citep[][]{Mathur2017,Uytterhoeven2011,Bruntt2012,MolendaZ2013,Mann2013,Sanchis-Ojeda2013,Furlan2018}, LAMOST \cite[including data from DR1 to DR5 and results of data analysis with different codes.][]{Luo2015,Frasca2016,Ho2017,Boeche2018,Luo2019,Xiang2019}, APOGEE \cite[{DR16} with ASCAP pipeline and APOKASC  studies.][]{Pinsonneault2018,Serenelli2017,Ahumada2020}.
When sources were common in multiple surveys, we adopted the value of [Fe/H] from the survey with the highest spectral resolution.
Furthermore, we selected only measurements of [Fe/H] with a reported precision better than 0.1 dex. 
For each metallicity measurement, we also kept the associated spectroscopic effective temperature.

Merging with the sample from section \ref{sec:periodsGaia} gives a sample of 7,914 stars with good quality Gaia DR2 data, rotation periods, and measurements of [Fe/H].  Table~\ref{tab:metal} lists the references for this final metallicity database, in order of selection priority and indicating the number of sources originating from each survey. For further reference, the full Gaia-\textit{Kepler}-Metallicity database is available online at the CDS. The final dataset used in this paper is flagged with \verb|MS_Cut=1| and its selection is described in the following sections.

Our choice to limit the metallicity precision to better than 0.1 dex is based on detailed comparisons of [Fe/H] values for sources that were in common in multiple surveys.  
We found that this precision cut worked best to eliminate (or reduce) the systematic differences (biases) observed between different surveys.
Furthermore, the analysis presented below searches for trends arising from relatively small changes in [Fe/H], which require this high precision.
Notably, our selection criteria excludes all measurements of [Fe/H] from the revised \textit{Kepler} Input Catalogue \citep[][and references therein]{Mathur2017}, which have typical precision between 0.15 and 0.3 dex. 

\begin{table}
\begin{center}
\caption{Spectroscopic metallicities.}
\begin{tabular}{ c c c c }
\hline
\hline
Selection & Sources & Resolution & Number of \\
order &     &     &  measurements  \\
\hline
1 & APOGEE ASPCAP DR16  & $\sim$22\;500 & { 1331} \\ 
&\citep{Ahumada2020} &      &   \\
2 & LAMOST DR5\footnotemark[1]  & $\sim$1\;800 & { 6570} \\
 & \citep{Luo2019} & & \\
 & \citep{Xiang2019} & & \\
3 & LAMOST DR1\footnotemark[2]& $\sim$ 1\;800 & { 2} \\
 & \citep{Ho2017} & & \\
4 & LAMOST DR1\footnotemark[3] & $\sim$ 1\;800 & { 11} \\
 & \citep{Boeche2018} & & \\
\hline
\end{tabular}
\label{tab:metal}
   \noindent\footnotemark[1]{with \verb|DD-Payne| (code labelled to APOGEE)}
   
   \noindent\footnotemark[2]{with \verb|The Cannon|}
   
   \noindent\footnotemark[3]{with \verb|SP_Ace|}
\end{center}
\end{table}

    \subsubsection{Trends in the full sample}

\begin{figure*}
\centering
\includegraphics[width=\textwidth]{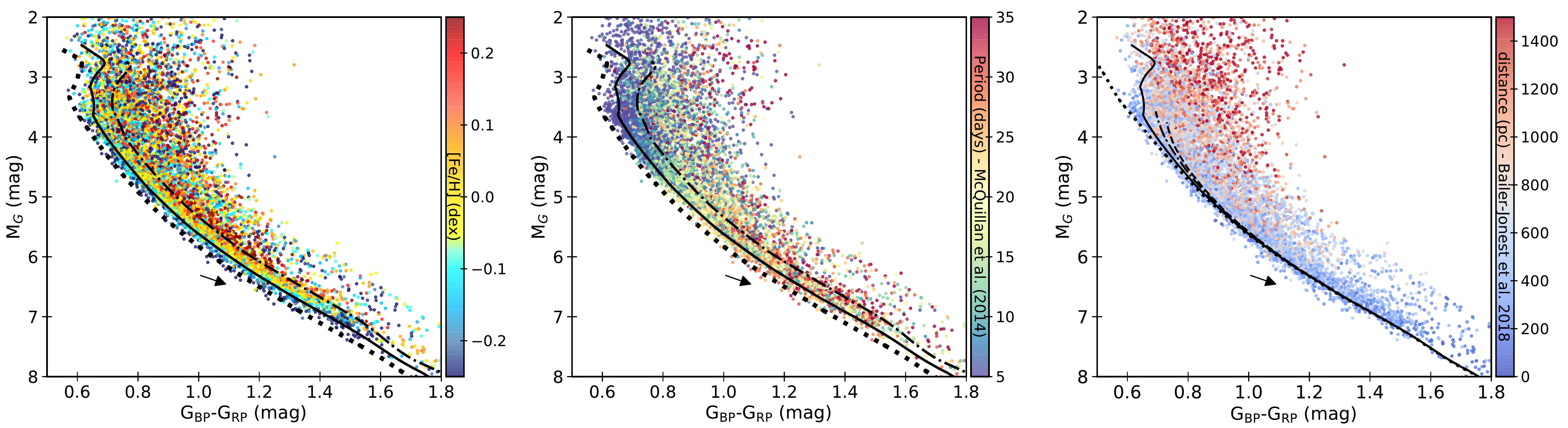}
\caption{\label{fig:CMD_init} Colour-Magnitude diagrams for 7914 stars with high quality Gaia DR2 data, rotation periods, and measurements of [Fe/H] (see \S~\ref{sec:fullsample}).
In the left and middle panels, the black lines show isochrones from \citet{Amard2019} with an age of 3 Gyr and for [Fe/H] = -0.3 (dotted line), [Fe/H] = 0 (continuous line), and [Fe/H] = +0.3 (dashed line). In the right panel, the lines show isochrones for [Fe/H] = 0 and ages of 1.5 Gyrs (dotted line), 3 Gyrs (continuous line), 4.5 (dashed-dot line) and 6 Gyrs (dashed line).
The black arrow shows the reddening vector for $A_V=0.213$ mag, which is the typical extinction expected for a star in the \textit{Kepler} Field at a distance of 1 kpc. 
Left: stars are coloured by their measured metallicity following the colour table in the right side of the panel. The colours saturates for stars with |[Fe/H]|>0.25 but all stars of the sample are displayed. 
Middle: stars are coloured by their rotation period in \citet{McQuillan2014}. 
Right: stars are coloured by their distance in Gaia-DR2 as in \citet{BailerJones2018}.}
\end{figure*}

Figure~ \ref{fig:CMD_init} displays the sample described above in a colour-magnitude diagram (CMD) with the relevant parameters of our dataset colour-mapped. 
From left to right, the panels are coloured by metallicity, rotation period and distance from our Sun, respectively.  
The three lines on the two left panels show theoretical isochrones from \citetalias{Amard2019} \citep[from][]{Amard2019}, with an age of 3 Gyr and [Fe/H] = -0.3, 0.0 and +0.3. 
In the left panel, although there is a large scatter, there is a clear overall gradient of metallicity perpendicular to the main sequence locus, indicating that redder/brighter stars are on average more metal-rich, as predicted in theoretical models.
In the central panel, the most obvious trend is a positive gradient of rotation period with magnitude, indicating that brighter stars are on average rotating faster than fainter stars.  
Among the brightest stars, there is a wide spread in rotation period, most likely because more slowly rotating sub-giant and giant stars are mixed with main sequence stars in this part of the CMD.
The right panel in Figure~\ref{fig:CMD_init} shows that the reddest stars in the upper CMD are on average further away, again indicating that this part of the CMD contains significant fractions of sub-giant and giant stars. On that same panel are shown four isochrones, all with solar metallicity, but with different ages, corresponding to 1.5, 3, 4.5 and 6 Gyr. It is clear that the different isochrones only cover a very limited part of the CMD. The spread of the entire main sequence can only be explained by assuming differences in chemical composition (also see the appendix)\footnote{Rotation can also increase the spread by changing the structure. However, since we only consider low-mass stars on their main sequence, the rotation rates are too small to cause any visible structural effects.}

    \subsection{The single, main sequence sample} \label{sec:singlemssample}

In order to focus our study on single, main-sequence stars, we make additional cuts to the data, based on distance and photometric binarity, described below.

\subsubsection{Distance cut: remove stars further than 1kpc} \label{sec:distancecut}

For subsequent analysis, we remove sources that are farther away than 1kpc, reducing the sample to 6,357 stars.
This cut achieves two things.
First, as mentioned above and highlighted by previous studies \citep[e.g.][]{Mann2012, Davenport2017}, a significant fraction of the brightest stars in the \textit{Kepler} field are giant and sub-giant stars. 
The spin evolution of such stars is strongly influenced by structural evolution and not only by the interaction of the large scale magnetic field and the stellar winds \citep{VSP13}. 
Due to this additional complexity, we restrict our analysis to main-sequence stars. 
Although \citetalias{McQuillan2014} already cleaned the sample from most evolved stars by removing low temperature-low gravity stars, the distance measurement that are now available thanks to Gaia allow for a more refined determination of the evolutionary status \citep[for a more detailed discussion see][]{Davenport2017}.
The restriction in distance naturally eliminates a significant fraction of evolved stars from the analysed sample.
Second, extinction (due to interstellar dust along the line of sight) changes a star position in the CMD, which can cause further scattering in the diagram (due to star-to-star variations) and possible trends, for example due to extinction being generally correlated with distance.
To evaluate the impact of extinction in the CMDs of Figure~\ref{fig:CMD_init}, we employed the 3D maps of interstellar reddening by \citet{Green2018}\footnote{\url{http://argonaut.skymaps.info/}}, which provides typical reddening values as a function of distance. Reddening values were transformed to extinction using $R_V=3.1$ and the appropriate relations provided by \citeauthor{Green2018}.
For the \textit{Kepler} field, the maps give an average extinction of $A_V= 0.213$  at 1 kpc, which produces a shift of $\Delta M_G=0.117$ mag and $\Delta BP-RP=0.096$ in the CMD.  
The reddening vector corresponding to this value of $A_V$ is shown by a black arrow on Figure~\ref{fig:CMD_init}, indicating that our distance cut also serves to reduce the extinction-related scatter, as well as possible trends with distance.
Finally, a byproduct of this distance cut is that the analysed sample does not contain any members of the known open clusters in the \textit{Kepler} Field, which are located at distances larger than 1.1 kpc.

\subsubsection{Removing photometric binaries}
\label{Sec:MS_selection}

A striking feature in the left panel of Figure~\ref{fig:CMD_init} is the gradient of metallicity along the Main Sequence in the CMD. It is clearly visible even in the full CMD with no attempt to account for giants, binary contamination or reddening applied. However, this gradient seems to repeat itself in a second sequence above the MS. This secondary sequence, located at about $2.5\log(2)\simeq 0.753$ mag above the MS, is generally understood to be due to unresolved (nearly) equal mass binary stars having the same colour, but double the flux of a single star. 
For the following analysis, it is important to remove the nearly-equal-mass binaries, primarily because this overlapping effect in the CMDs will confuse the search for trends in the data.

The common procedure to identify and separate single stars from unresolved (nearly) equal-mass binaries, is to place an isochrone of a chosen age and metallicity inside the CMD, and select as single MS stars the sources that are up to  $\Delta M_G\simeq-(0.376+\sigma_{M_G})$ brighter than the isochrone, and down $\Delta M_G\simeq+\sigma_{M_G}$ fainter, where $\sigma_{M_G}$ is the typical uncertainty in $M_G$.  
The threshold offset in magnitude of $-0.376$ is half of the magnitude difference for an equal-mass companion, so this cut is meant to exclude most nearly-equal-mass binary stars, determined from photometry alone.  However, given the metallicity trend seen in the left panel of Figure~ \ref{fig:CMD_init}, a MS selection that uses a single metallicity isochrone would result in including metal-poor, equal-mass binaries and excluding the most metal-rich, single stars. For example, considering the 3 Gyr isochrones shown in Figure~ \ref{fig:CMD_init}, a star with colour $BP-RP=1$\footnote{For a readability purpose, $BP$ and $BP$ represent $G_{RP}$ and $G_{BP}$, respectively.} in the [Fe/H]$=-0.3$ isochrone will have a $\Delta M_G=+0.49$ mag compared to a star with the same colour in the [Fe/H]$=+0.3$ isochrone.  Therefore, we apply a metallicity-dependent single-star selection. We start by binning our sample using the same steps of [Fe/H] for which the \citetalias{Amard2019} isochrones are available, which defines 6 metallicity bins following the limiting values of [Fe/H]$=-1.0,-0.5,-0.3,-0.15,0.0,+0.15,+0.3$ dex. 
Within each metallicity bin, we select as single-MS stars the sources that lay between a 5 Gyr isochrone for the high end of the metallicity bin, shifted by $\Delta M_G=-0.376+\sigma_{M_G}$ and $\Delta (BP-RP)=+\sigma_{BP-RP}$, and a 1 Gyr isochrone for the lower end of the metallicity bin, shifted by $\Delta M_G=+\sigma_{M_G}$ and $\Delta (BP-RP)=-\sigma_{BP-RP}$.  We adopt $\sigma_{M_G}=0.060$ mag from the maximum photometric uncertainty expected in the Gaia DR2 photometry, taking into account the effect of the parallax uncertainty in the estimation of the absolute magnitudes, and $\sigma_{BP-RP}=0.014$ mag from the maximum uncertainty expected for the BP-RP colour. 
The choice to use isochrones with different ages at the lower and upper bound of each bin makes the bins slightly wider than if using a single age, and this takes into account the wide range of ages expected for the \textit{Kepler} field stars.
This procedure reduces the full dataset to a sample of 4060 stars, which should have a significantly reduced contamination by sub-giants and (photometric, nearly-equal-mass) binary stars. 

Finally, we also checked our sample against the sources in the Third Revision of the Kepler Eclipsing Binary Catalog\footnote{\url{http://keplerebs.villanova.edu/}} \citep{Kirk2016,Abdul-Masih2016} and we eliminated 5 extra stars that are listed as eclipsing binaries by the third revision of the catalog. The final sample of 4055 stars is shown in CMDs in Figure~ \ref{fig:CMD_MS}, coloured by metallicity (left panel) and rotation period middle panel), and stellar masses (right panel) with zoomed insets to highlight the trends discussed below. 

Note that \citet[][]{Berger2018,Berger2020} recently did a thorough job at characterising the Gaia-\textit{Kepler} sample using individual extinction and re-computed distances. In particular, they estimate the age and evolutionary stage of each star as well as whether they are in a binary system or not. A cross-checking of our samples with \cite{Berger2018} reveals that only 15 stars of our final sample were flagged as evolved or binary. This small difference likely comes from our metalicity-dependent binary sequence definition. \citet{Berger2018} define theirs with a shifted solar metallicity MESA isochrone, while in our case, we use the information on metallicity of each star (or stellar system) combined with STAREVOL isochrones at the closest metallicity to define the binary sequence. The distances we got from \cite{BailerJones2018} are extremely close to \cite{Berger2020} as a difference only becomes noticeable beyond 2500 pc.

\subsubsection{Mass distribution}
\label{sec:mass_determ}
We use a maximum-likelihood interpolation tool adapted from \citet{Valle2014} to derive from the model grid the mass of each star of the MS sample described in Sec~\ref{Sec:MS_selection}. The tool compares to the theoretical tracks the Gaia M$_\textrm{G}$ absolute magnitude and $G_\textrm{BP}$-$G_\textrm{RP}$ colour as well as the spectroscopic values of the metallicity and the corresponding effective temperature of the observed stars, and provides us with a mass (with an error of about 3 to 5 percent). 
Note that for the most massive stars of our sample (typical F-stars), chemical transport through atomic diffusion may happen and change the stellar surface parameters \citep[\textit{e.g.}][]{Turcotte1998}. The models we are using do not include it and the mass determination may then be affected. Fortunately, the effects of rotation-induced mixing have been shown to be dominant below 1.4 M$_\odot$ \citep{Deal2020} and these are included by the \citetalias{Amard2019} models of the present grid. The mass distribution of our sample is presented at the bottom panel of Figure~\ref{fig:P_FeH_dist}, and displayed on a CMD in the right panel of Figure~ \ref{fig:CMD_MS}.

   \subsubsection{Trends in the single, main-sequence sample}

\begin{figure*}
\centering
\includegraphics[width=\textwidth]{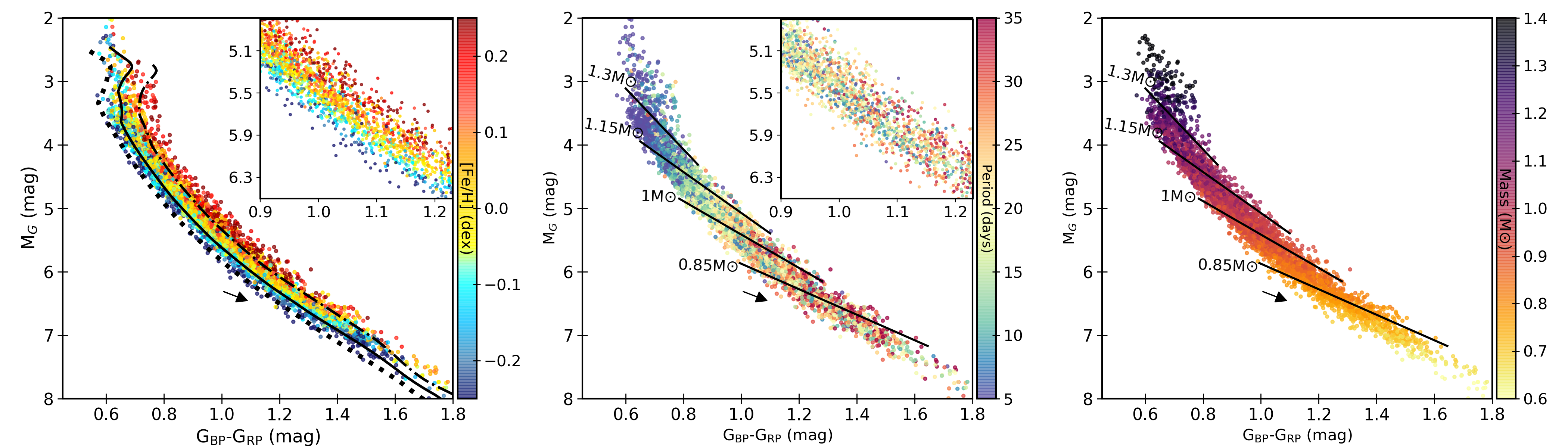}
\caption{\label{fig:CMD_MS}
\textbf{Left and middle:} Same diagrams as the left two panels of Figure~\ref{fig:CMD_init}, but showing the 4055 single, main-sequence stars selected in section~\ref{sec:singlemssample}.
In the left panel, the black lines are the same as the ones on the two left panels of fig.~\ref{fig:CMD_init}.
In the middle and right panels, the black lines show positions of stars with approximately constant mass 
(1.3, 1.15, 1.0 and 0.85 M$_\odot$), as indicated. In the left and central panel the inset figure highlights a subset of data to show various features described in the text. The right panel shows the same CMD coloured with the mass of each star obtained following the method described in \S~\ref{sec:mass_determ}.}
\end{figure*}

\begin{figure}
\centering
\includegraphics[width=0.95\columnwidth]{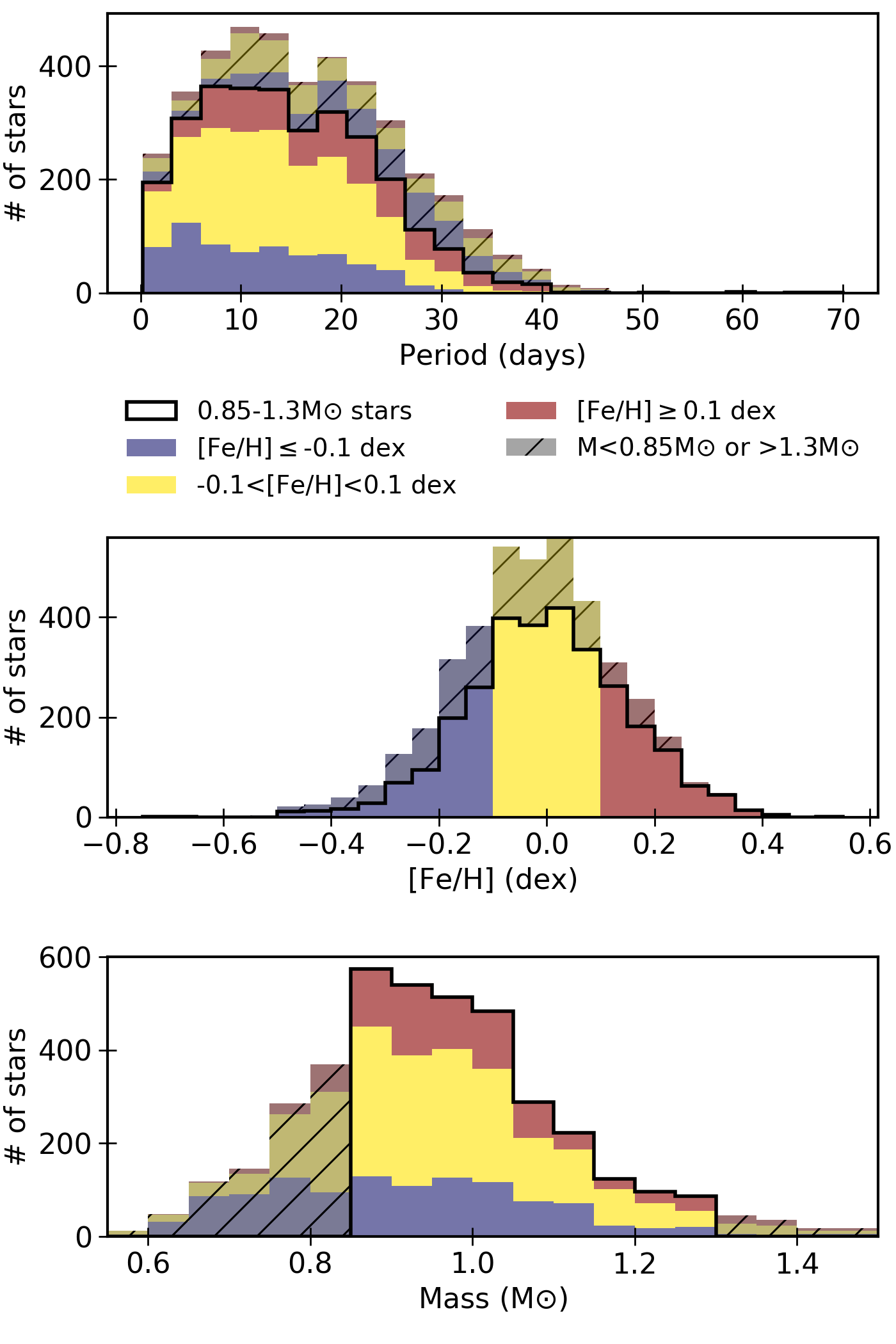}
\caption{\label{fig:P_FeH_dist} Histograms presenting the rotation period (top),  [Fe/H] (middle), and mass (bottom) distributions of the "main sequence" sample. The colours indicate the three metallicity bins indicated in the legend, blue for low [Fe/H], yellow for mid [Fe/H] and red for high [Fe/H]. The greyed parts show the stars which have been removed from the main sequence sample due to the mass cuts defined in \S~\ref{sec:massmetals}.}
\end{figure}

The left panel of Figure~\ref{fig:CMD_MS} shows a strong gradient of metallicity in a direction perpendicular to the main sequence locus, and the magnitude and direction of the effect is similar to that predicted by stellar models (e.g., following the isochrones on the Figure).
The clarity of this correlation between metallicity and CMD position is due to the high precision afforded by high resolution spectroscopy (for metallicities) and the distances and magnitudes from Gaia, combined with our selection of single, main-sequence stars.
This trend is well understood from stellar structure theory \citep[\textit{e.g.}][]{Kippenhahn2012}.  
An increase in metallicity of a star makes it globally more opaque to radiation and so inhibits the transport by photons in the stellar interior. 
As a consequence, for a given mass, less energy reaches the stellar surface, and the star appears less luminous. 
The radiative gradient in the star is also increased with the opacity and leads to a deeper convective envelope and a lower effective temperature. 

The middle panel of Figure~\ref{fig:CMD_MS} highlights the rotation distribution for our main sequence sample, in a similar way as Figure~1 in \citet{DavenportCovey2018}. The rotation-mass trend, well established by the theory today \citep[\textit{e.g.}][]{Matt2015}, is here clearly visible---brighter and hotter stars are essentially fast rotators and the lower end of the main sequence is mostly constituted with slow rotating stars. 
The spin-down of low-mass stars on the main sequence is highly dependant on the ability of the star to generate a large scale magnetic field through a convective dynamo mechanism. 
Lower mass stars have globally deeper convective envelopes and slower convective motions. Consequently, at a given rotation period, they are expected to have a stronger and larger scale magnetic field and thus spin down more efficiently \citep[See \textit{e.g.}][]{BrunBrowning2017}.

\citetalias{DavenportCovey2018} highlighted an additional trend in the rotation period that is perpendicular to the main-sequence locus.  
Specifically, they noted that redder/brighter stars are on average slower rotators than bluer/fainter stars, which is visible in the middle panel of Figure~\ref{fig:CMD_MS} as a weak colour gradient orthogonal to the main sequence, most clear in the inset panel.
The trend in our dataset appears somewhat weaker as in that of \citetalias{DavenportCovey2018}, which might be due to our dataset being smaller or to \citetalias{DavenportCovey2018} highlighting a fainter/redder section of the CMD (where we have much fewer numbers).
Regardless, the same trend is visible in our data, and we further analyse this feature in the next section.

\section{Tracing the metallicity influence on rotational evolution}
\label{sec:Zinfluence}

  \subsection{Disentangling mass and metallicity} \label{sec:massmetals}

    The most obvious trends with rotation that are visible in Figures~\ref{fig:CMD_init} and \ref{fig:CMD_MS} is the strong correlation between the rotation period and the position along the main sequence locus.  In order to better analyse the trends in the direction orthogonal to the main sequence, it is useful to first extract the effect of stellar mass on the distribution.
    The solid lines in the middle and right panels of Figure~\ref{fig:CMD_MS} show the locations of stars with approximately constant mass, with values of 0.85, 1.1, 1.15 and 1.3 M$_\odot$.
    These are linear fits to the CMD locations of stellar models of a given mass at all available metallicities, using the 3 Gyr isochrones.
    It is clear that, at a fixed mass, more metal rich stars are dimmer and redder, and they rotate more slowly on average, than metal poor stars.
    
    To further explore the trends, we cut the sample to exclude stars with masses greater than 1.3M$_\odot$ and less than 0.85M$\odot$ (2931 stars).  
    This cut is to avoid the incompleteness in our sample at both ends, and it also further removes remaining subgiants from the sample at the high-mass end. 
    This final sample properties are shown in the histograms of Figure~\ref{fig:P_FeH_dist} presenting the [Fe/H], rotation period, and mass distributions, before and after the mass cut.
    The metallicity distribution peaks around solar metallicity with the tails decreasing regularly on each side, down to [Fe/H]=-0.5 and +0.4. The rotation periods peak around 10-14 days with the numbers quickly dropping on the smaller periods side and more slowly towards longer periods. The stars removed from the mass cut (dashed on the histograms) are distributed fairly uniformly in metallicity but are biased towards fast rotators. We removed the stars beyond 1.3 M$_\odot$ which are expected to be very close or beyond the Kraft break \citep{Kraft1967}. These are not spun down on the main sequence because of their extremely thin convective envelope and thus their spin evolution change according to different processes. We also removed stars below 0.85 M$_\odot$ because of the inhomogeneity of the data set. As it can be seen in the bottom panel of figure~\ref{fig:P_FeH_dist}, very few high metallicity stars and, on the contrary, a lot of low-metallicity stars are present. The latter are more globally brighter, explaining why low-metallicity, low-mass stars are more easily observed than their metal-rich counterpart.

\begin{figure*}
\centering
\includegraphics[width=\textwidth]{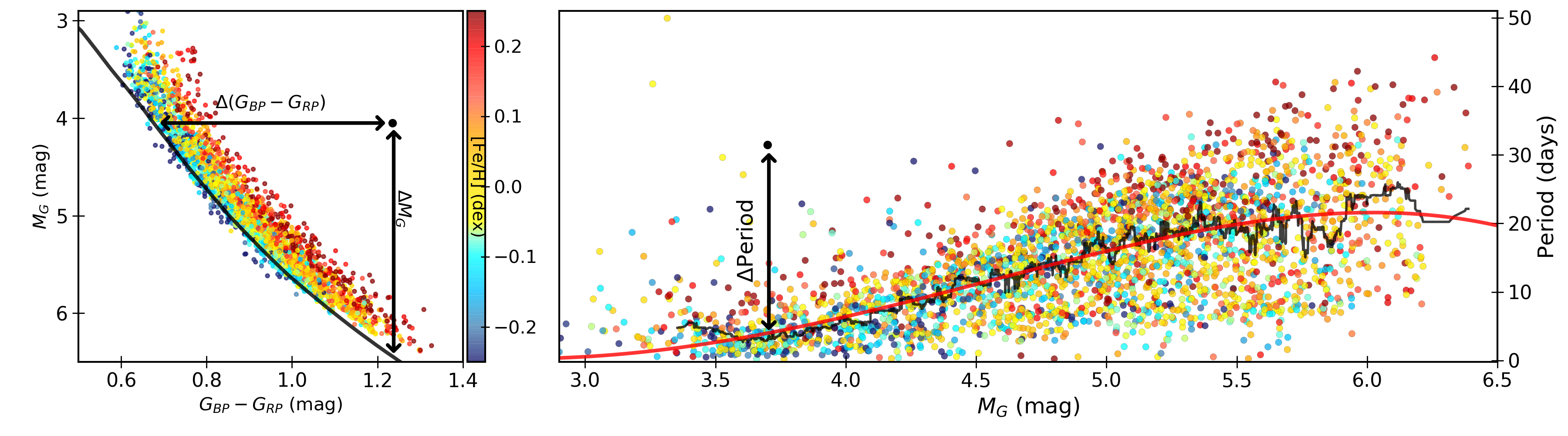}
\caption{\label{fig:Per_Mag_runmed}
\textbf{Left:} Colour-magnitude diagram colour-coded with [Fe/H] with the same sample as Figure~\ref{fig:CMD_MS}, but trimmed to only include stars with masses between 0.85 and 1.3~$M_\odot$. The black line shows a 1~Gyr solar-metallicity isochrone from \citet{Amard2019}. Compared to the 3~Gyr isochrone shown in previous Figures, this younger one allows to give a unique value to $\Delta M_G$ and $\Delta(\textrm{G}_\textrm{BP}-\textrm{G}_\textrm{RP})$. \textbf{Right :} Rotation periods as a function of Gaia Magnitude for the same data and colour code as the left panel. 
The red line shows a polynomial fit of order 4 to the running-median (black line). Black arrows and annotations illustrate how we compute the difference quantities that are plotted in Figure~\ref{fig:Delta_DeltaPer}, for each star (black point is for illustration only). 
}
\end{figure*}

    The sample is then displayed in a CMD on the left panel of Fig~\ref{fig:Per_Mag_runmed}.
    In order to extract the variations in rotation that may be due to chemical composition, one needs to define a referential independent of other parameters. As a first step, we want to show an empirical comparison based on observational properties that can be compared with previous work \citep{DavenportCovey2018}. To do so, we realise an analytical fit to the observed rotation period distribution as a function of the stellar magnitude as shown on the right panel of Fig.~\ref{fig:Per_Mag_runmed}. 
    The rotation period increases with magnitude as it would be expected of lower mass stars. The magnitude can hereby be seen as a empirical proxy for stellar mass \citep{Douglas2014}\footnote{Note however, that it really is a very rough approximation as it can be seen from the iso-mass displayed in the left panels of fig.~\ref{fig:CMD_MS}.}.
    Already in this diagram we see the presence of a metal-rich (red) sequence at slow rotation rate for all magnitude, indicating potentially a slower rotation by high metallicity stars.
    Note also that, as a consequence of the mass cut we operated in the previous section, the metallicity distribution is not the same over the whole range, for example high magnitudes are dominated by Solar-to-high metallicity stars. We keep in mind that this may affect our conclusion and will try to account for it in \S~\ref{sec:comp_synth} by using the right couple mass-metallicity.

    To derive the fit used as a referential, we use a window of 50 data points to calculate the moving-median of the rotation period distribution as a function of magnitude (black solid line on right panel of Fig.~\ref{fig:Per_Mag_runmed}).
    We then derive an analytical function for the magnitude-dependence by performing a 4th order polynomial fit to the moving-median of the data, shown as a red curve in the right plot of Figure~\ref{fig:Per_Mag_runmed}.
    For each star of the sample, we then estimated a $\Delta P_{\textrm{rot}}$ defined as the difference between the star's period and the period given by the empirical relation using the star's magnitude.
    A positive $\Delta P_{\textrm{rot}}$ thus means the star is rotating slower than the median of the sample at this magnitude. 

\begin{figure*}
\includegraphics[width=0.4\textwidth]{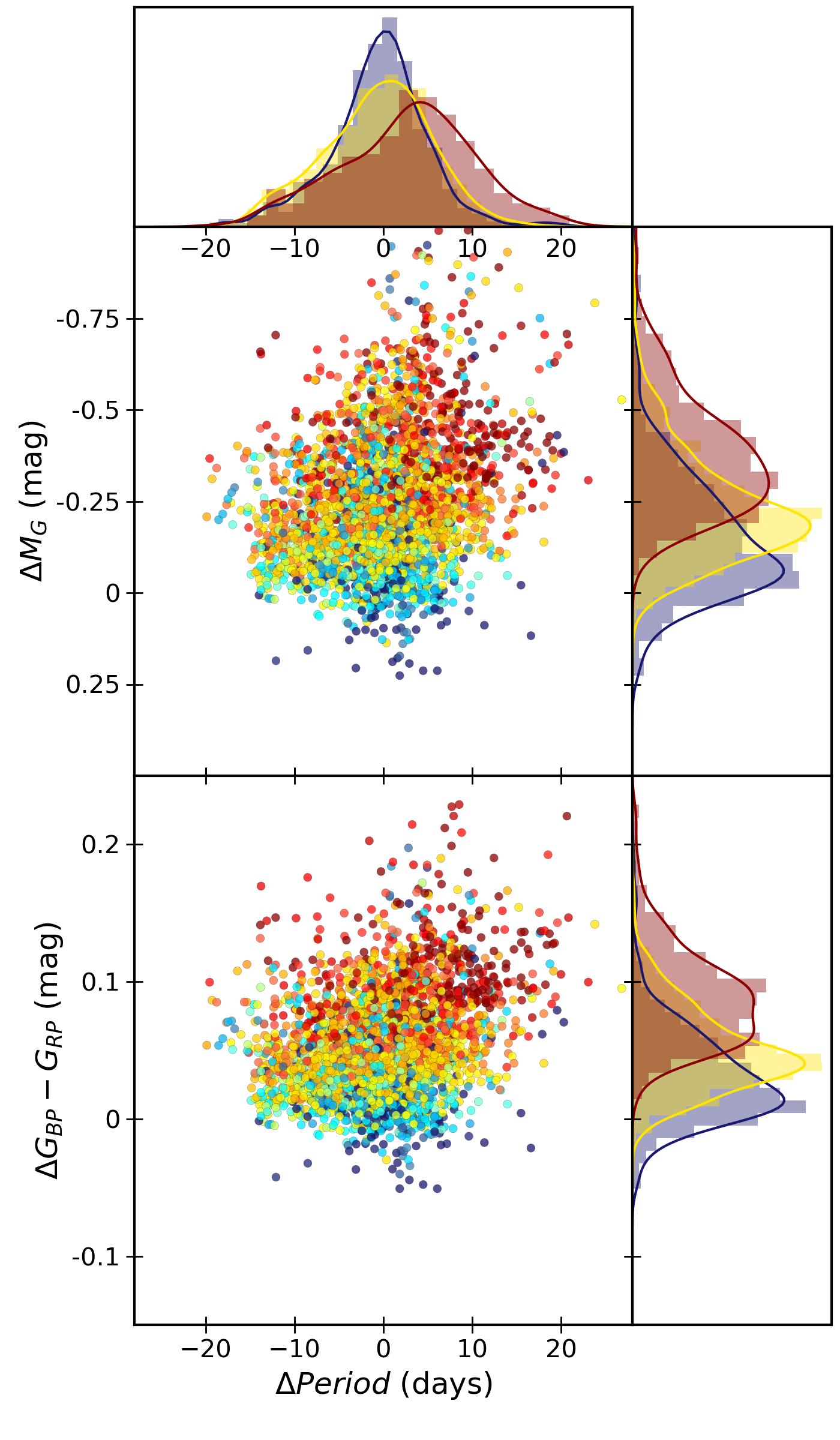}
\includegraphics[width=0.4\textwidth]{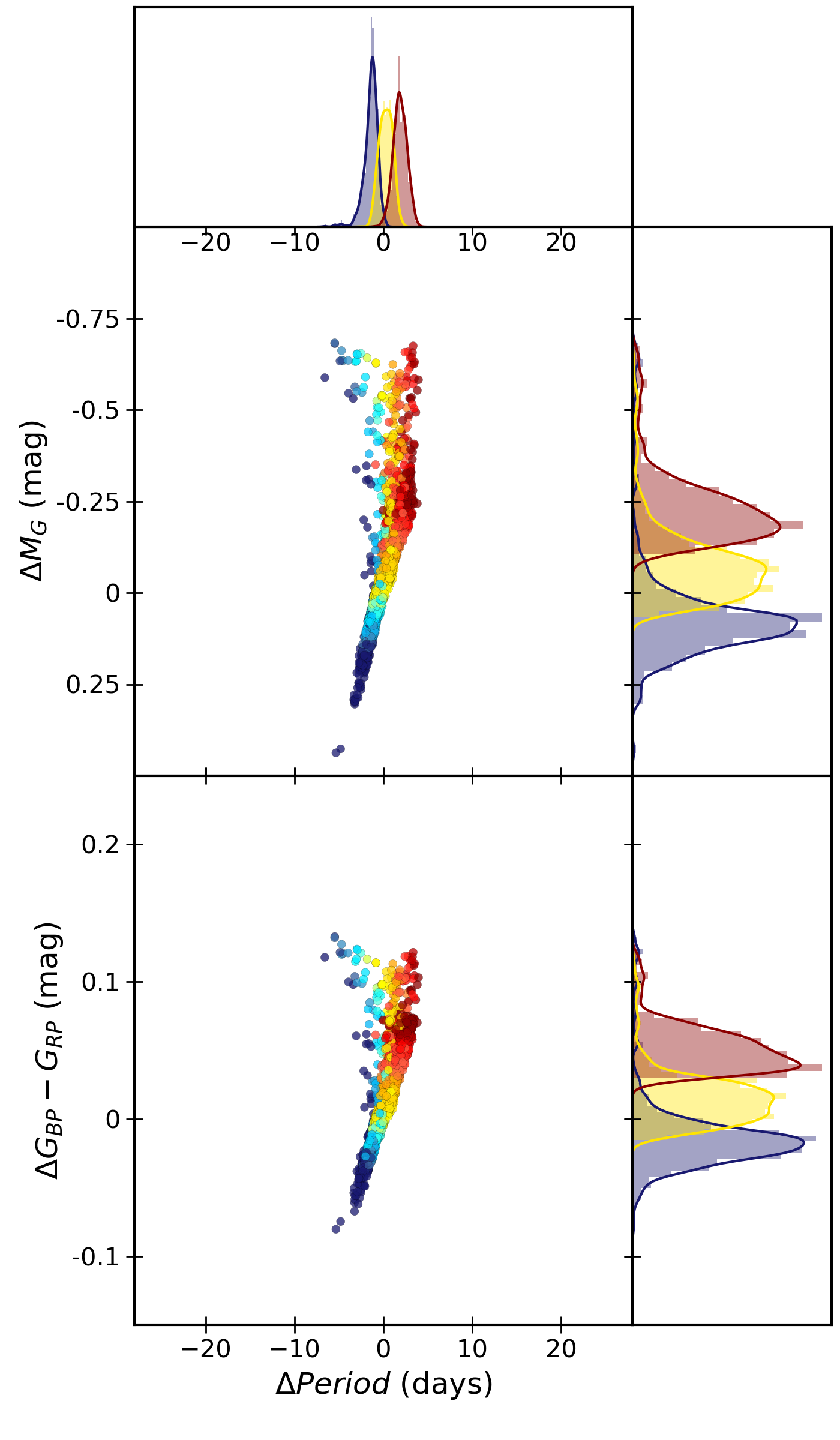}
\caption{\label{fig:Delta_DeltaPer}
Visualisation of the delta quantities defined in Figure~\ref{fig:Per_Mag_runmed} for the observed sample (left set of panels) and the 3-Gyr-old synthetic sample (right set of panels). Both samples have been limited to stars with a mass between 0.85 and 1.3 M$_\odot$.
\textbf{Top.} Difference in magnitude $M_G$ from a 1 Gyr solar metallicity isochrone as a function of the difference in rotation period from the running median of each sample for the observed (left) and modelled (right) sample. 
\textbf{Bottom.} same as the top panels but showing the difference in colour $G_{BP}-G_{RP}$.
The colour indicates the metallicity of each star with the same colour scale as in the left panel of fig.~\ref{fig:CMD_MS}. The histograms on the right side of each panel show the normalised distribution (indicated by the KDE) of the three metallicity bins discussed in \S~\ref{sec:Zinfluence}: metal-poor (blue, [Fe/H]<-0.1); solar-metallicity (yellow, -0.1<[Fe/H]<0.1); and metal-rich (red, [Fe/H]> 0.1). 
}

\end{figure*}

    The results are presented as a function of the offset from a solar metallicity 1 Gyr isochrone, in magnitude (top) and colour (bottom) on the left side of Figure~\ref{fig:Delta_DeltaPer}. These offsets are illustrated visually in the left panel of Figure~\ref{fig:Per_Mag_runmed}. Positive offset values correspond to stars that lie above the isochrone.
    The stars are still colour-coded with metallicity, metal-rich stars are clearly redder/brighter stars than their metal-poor counter-part. We split the sample in three subgroups with [Fe/H]<-0.1 (692 stars), -0.1<[Fe/H]<0.1 (1535 stars), and [Fe/H]>0.1 (704 stars). 
    Their kernel density estimations (KDE) plotted on the side help to show the correlation between $\Delta P_{\textrm{rot}}$ and the offset in colour.  The $\Delta P_{\textrm{rot}}$ distribution shown above the diagrams indicates a shift of the metal-rich subgroup towards rotation periods longer than the median, confirming the link with metallicity of the feature observed by \citetalias{DavenportCovey2018}. 
    Brighter and redder stars are metal-rich and, on average, slower rotators. Now, what is clearly visible on this diagram is the trend with metallicity.
    Stars with a $\Delta P_{\textrm{rot}}$ > 0 (\textit{i.e.}, rotating slower than the median) are cooler and brighter, as seen in \citetalias{DavenportCovey2018}, but are also systematically more metal-rich.

\subsection{Metallicity vs rotation period} \label{sec:periodmetals}

\begin{figure}
\centering
\includegraphics[width=0.48\linewidth]{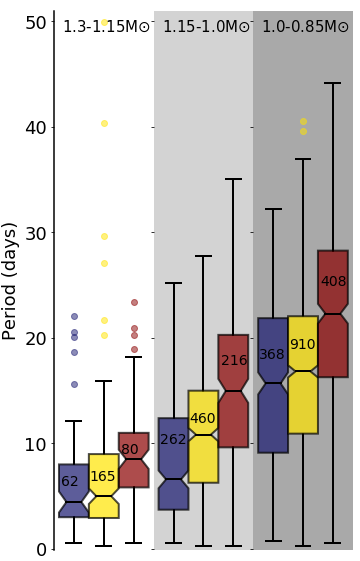}
\includegraphics[width=0.48\linewidth]{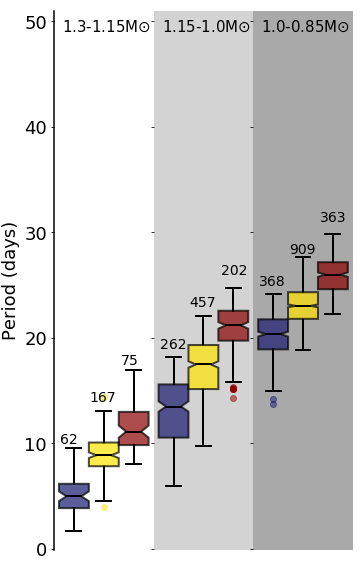}
\caption{Statistical distributions of rotation periods for different masses and metallicities.  \textbf{Left panel:} distributions observed in the \textit{Kepler} field sample.  \textbf{Right panel:} distributions predicted by a 3-Gyr-old synthetic sample.  In each panel, from left to right, box plots display the observed rotation period distributions in the $[1.3; 1.15]$, $[1.15; 1.0]$, and $[1.0; 0.85]$ M$_\odot$ mass bins, respectively, for the low (blue), solar (yellow) and high (red) metallicity bins. The median is indicated by the notch, the first and third quartiles define the vertical extent of the box, the whiskers' extremities are located by default at one and a half times the interquartile range (or reaching the last data value, whichever comes first), and the small circles show all outlying points. The number of sources in each box-plot is shown inside the box.} 
\label{fig:Boxplot}
\end{figure}

In order to further explore correlations between metallicity and rotation, we split the sample in the three metallicity bins described above.
In each sub-sample, we also split the stars into three mass bins, delimited by 1.3, 1.15, 1.0 and 0.85 M$\odot$ shown in the right panel of Figure~\ref{fig:CMD_MS}.
The left panel of Fig.~\ref{fig:Boxplot} displays the statistical properties of the rotation period distribution of the nine sub-samples in a box plot.
Globally, the median rotation period increases as we go towards lower masses and as the metallicity of the sub-sample increases. 
Although the amplitude of the effect is not exactly the same, in each mass bin, the high metallicity bin (red) shows the slowest rotation rate, and the metal poor sub-samples (blue) are faster rotators. This result is so far in agreement with \citet{AmardMatt2020}, at least qualitatively. In the next section we realise a synthetic sample with rotating stellar models to test the robustness of this conclusion.
Overall, it is clear that the median rotation period is correlated with metallicity, and that this correlation is most clear at a fixed stellar mass.

\section{Comparison to rotation evolution models}
\label{sec:comp_synth}

In order to better understand the correlation between metallicity and rotation period, we created a synthetic distribution based on one existing model.
We compare the properties of the observational sample with a synthetic rotation distribution from the grid by \citet{Amard2019} which included a consistent treatment of rotation along the evolution and has been calibrated on young open clusters.  The grid uses a physics suitable for low-mass stars and was computed with metallicities and $\alpha$-element enhancement corresponding to the galactic thin disc, for more information see the original paper.

\subsection{Properties of the synthetic sample}
\label{sec:prop_synth_sample}

For each star, using its metallicity and fit mass (see sect.~\ref{sec:mass_determ}), we determine the predicted properties by interpolating from the model grid at a fixed age of 3~Gyr.
Specifically, we retrieve the colour, magnitude, and rotation period given by the models. The use of a single age population is far from realistic but allows us to avoid any bias in terms of possible age distribution and to show an extreme case.
Another possibility would have been to use the ages fit from the model grid, however the errors in age are quite large.  
Furthermore, since the rotation period is very sensitive to the age, this process would introduce a bias.
Thus, we fixed the model age at 3~Gyr, for illustrative purposes.
In the appendix, we show how the synthetic distribution looks at different ages, illustrating that the main effect of an age spread is to increase the spread in rotation periods at a given mass and metallicity.
As for the initial rotation period of each synthetic star, since we do not have any constraints, we randomly chose a value between 1.6 and 9 days, which is roughly the range of periods covered in very young clusters. 
At 3~Gyr old, all the stars in our sample have converged towards the same rotational evolution, so variations in the initial period have very little effect.

The right panels of Figures~\ref{fig:Delta_DeltaPer} and \ref{fig:Boxplot} show the properties of the synthetic sample, in the same format as for the observational data.
The synthetic sample underwent the exact same trimming as the observed data set, described in sections~\ref{Sec:MS_selection} and \ref{sec:massmetals} for Figure~\ref{fig:Delta_DeltaPer} and section~\ref{sec:periodmetals} for Figure~\ref{fig:Boxplot}.
In order to make the right panels of Figure~\ref{fig:Delta_DeltaPer} we determined a fourth-order fit to the synthetic rotation distribution in the P$_\textrm{rot}$-M$_G$ diagram and computed the offset $\Delta$Period from it.

\subsection{Comparison to observed sample}

In the right panel of Figure~\ref{fig:Delta_DeltaPer}, the synthetic sample shows a very clear trend that low-metallicity models are bluer, fainter, and slightly faster rotators than metal rich models.
The trend in the synthetic sample is much more clear (showing less scatter) than the observed distribution, reflecting the fact that the synthetic sample assumes a single age.
The \textit{Kepler} field stars are expected to have a wide age distribution with a peak around 3-4~Gyr for the thin disk population \citep{Miglio2020}.
This spread in age should naturally produce a wide scatter in delta period (and a somewhat smaller scatter in delta mag and delta colour).
Indeed, while the observed stars in the high-metallicity bin clearly show a slower average spin rate, the solar- and low-metallicity bins are too broad to be obviously distinguished.
However, in spite of the scatter, it appears that the synthetic sample reproduces the general trend of delta period with metallicity that goes in the same direction as observed and has the same magnitude (i.e., delta period is predicted to be +/- a few days for high/low metallicity stars).

On the right panel of Figure~\ref{fig:Boxplot}, the median rotation periods of the synthetic sample shows the same overall trend with mass and metallicity as the observed sample.
The synthetic median periods are not identical to the observed values, and the synthetic period distributions are much narrower, but these are primarily due to the assumption of a single age for the synthetic distribution (see appendix).
The most interesting comparison is in the dependence of median period on metallicity.
The synthetic sample predicts a monotonic dependence of rotation period on metallicity, with a difference of a few days between each metallicity bin.

The existence of a large age spread in the sample should confuse these trends, so the effect of metallicity on stellar rotation rate must be strong in order for it to be still visible in a population such as the \textit{Kepler} field stars.
The strength of the observed trend appears to be of a similar order as that predicted by modern theoretical models (such as that used for the synthetic sample).

\section{Discussion}
\label{sec:discussion}

  We showed that the observed correlation between metallicity and rotation period might be explained by metallicity-dependent rotational evolution.  
  However, the observed trend could in principle be explained instead by metallicity-dependent age distributions or metallicity-dependent detection biases.
  Here we discuss all three possibilities.

  \subsection{A dependence of stellar wind torque on metallicity} \label{sec:braking}

    How exactly the metallicity can affect the rotational evolution is still uncertain. 
    It is not likely due to a difference in initial rotation rates (at birth), since the stars of our samples are on the main sequence, and most of their rotation rates are expected to have converged, such that their initial conditions have effectively been erased.
    Thus, the observed trends suggest that stars with the same mass and rotation rate will have a spin-down torque that depends on metallicity, with more metal rich stars having a stronger torque.
    
    In section~\ref{sec:comp_synth}, we used models that adopt one particular prescription for stellar wind torques \citep{Matt2015,Amard2019}, in which the metallicity dependence arises primarily from the strong dependence of the torque on stellar Rossby number, which is itself strongly influenced by metallicity \citep{AmardMatt2020}.
The mass-dependence of rotation period in the synthetic sample arises from the assumptions in the physical models, which are, at least partially, tuned to fit the mass-dependence in observed single-age populations (and even then, the models do not perfectly fit the observed trends).
Thus, the general agreement of the models to the observed mass-dependence is not surprising (i.e., the rotation distributions in field stars broadly follows the same mass dependence observed in single-age populations, to which the models have been tuned).
On the other hand, the metallicity-dependence in the models is a true prediction, in a sense that the formulation for the torque was derived without directly considering metallicity effects and in the absence of observational information about how metallicity might affect spin down.
It is promising that the observed trend goes in the same direction and is of the same magnitude as predicted.

Note also that the angular momentum loss prescription used in the \citetalias{Amard2019} models does not account for a reduced braking beyond a given Rossby number, as was originally demonstrated by \cite{VanSaders2016}, and tested with more or less success by several authors since then \citep[][Hall et al. subm]{Metcalfe2017,Kitchatinov2017,MetcalfeEgeland2019,VanSaders2019,LorenzoOliveira2019,Booth2020}. Conveniently, none of the stars  of our sample with an efficient torque on the main sequence have reached this Rossby number by 3 Gyrs. The weakened breaking would only affect the hottest stars of the sample, which already hardly spin down on the MS. 

While other description of the angular momentum loss may provide different results, the \citetalias{Amard2019} models uses the torque by \citet{Matt2015}. The latter has an extra mass dependence but no extra variation with the surface properties, which may change the stellar wind properties and so the torque. Its simplicity makes its strength, some other prescriptions may be more accurate in the modelling of the torque but require more complex physical inputs that are not readily available from evolution models \citep{Reville2015,PantolmosMatt2017,FinleyMatt2017, FinleyMatt2018}.
Nevertheless, there are some other wind prescriptions in the literature which only require relatively simple inputs, and that can be obtained from an evolution models \citep{VSP13,VanSaders2016}. 

\citet{AmardMatt2020} also showed that depending on the braking law that is used, low-mass stars can converge toward a very similar rotation evolution once they are in the unsaturated regime. 
The sample and our conclusion do not allow to favour one torque to another. Even if it did, the very large discrepancy of each low-mass sub-sample and the numerous biases of the global sample lead us to be cautious with such claims.

So far, the above discussion only considers the effects of metallicity on stellar wind torques via its influence on stellar structure and Rossby number.  However, metallicity is also expected to affect coronal heating and mass-loss rate \citep{Suzuki2018, WashinoueSuzuki2019} and other aspects of stellar activity \citep{Witzke2018}.  We are not aware of published work on the effect of composition on the dynamo process, although it has been observed to affect the latitudinal differential rotation \citep{Karoff2018}, which is strongly related to the underlying dynamo process \citep{Brun2017,BrunBrowning2017}. 
Therefore, it is likely that all of the stellar wind torque prescriptions currently being used are still missing some physics to properly include the effects of metallicity.
Further modelling and comparison with future data sets, including metallicity information, may help to further constrain our understanding of stellar wind torques, and underlying processes, such as coronal physics and stellar dynamos.

  \subsection{Age distributions correlated with metallicity}

    As an alternative explanation, the correlation between rotation period and metallicity could in principle be entirely explained by a correlation between metallicity and age distributions. 
    Both from models and from observations of single-aged populations, we know that the distribution of rotation rates is very sensitive to age.
    In particular, if the metal rich stars in our sample are older on average than metal poor stars, it could explain the observed trend (without any metallicity dependence in braking torques).
    
    \citet{DavenportCovey2018} pointed out that a positive correlation between age and metallicity goes against expectations from global galactic chemical evolution, which predicts an increase of the metallicity as time goes on. 
    On the other hand, even though globally the metal content of the galaxy is increasing, stars with lower metallicity than the Sun are still being formed today, as can be seen in some young open clusters. 
    It is commonly accepted in the literature that the age-metallicity relation is relatively flat, with a large scatter in metallicity for all ages younger than about 8 to 12 Gyr \citep[e.g.][]{Haywood2013,Lin2020}. Thus, a star's metallicity cannot generally be used as an indicator of its age. Of course, the specific star formation history (or importantly the age-metallicity relationship) of the stars in our sample could deviate from global galactic trends. This explanation seems unlikely, but in the absence of any age information, it cannot be ruled out. 
    For example, \citet{Nissen2020} recently showed the existence of two sequence in the age-metallicity relation.  A first population of metal-rich stars would have been created about 7 Gyr ago and a second star-forming sequence over the last 6 Gyr with increasing metallicities from -0.3 dex to +0.2 dex today.  If the same star formation history is applicable to the Kepler field, this could at least partially explain the metallicity trend on the fainter and slowly rotating part of the CMD.

  \subsection{Dependence of rotation period detection on metallicity} \label{sec:detectionbias}

    As another alternative, the correlation between rotation period and metallicity could in principle be entirely explained by a correlation between metallicity and the detectability of periods.
    Not all stars have detected periods. 
    First, because of limited observing time it is generally harder to detect rotation periods in slow rotators than in fast rotators. 
    Second, for a given T$_\textrm{eff}$ range, the amplitude of variability due to rotation generally decreases as rotation rates get slower \citep{McQuillan2014}. 
    Consequently, the fraction of stars with detected rotation periods is expected to drop with longer periods, skewing most observed samples toward shorter average periods.
    If, for example, the amplitude of variability is less for metal poor stars, then the population of detected periods will be more skewed to shorter periods, relative to the population detected in high metallicity stars, possibly explaining the observed trend.
    
    The amplitude of variability could be physically explained by the link between stellar Rossby number and magnetic activity. The periodic photometric signal of rotation is generally understood to be caused by inhomogeneously distributed dark and/or bright spots on stars' surfaces, due to surface magnetic activity.
    The latter was shown to be correlated with the ratio of the rotation period and the convective turnover timescale---the higher the stellar Rossby number, the smaller the magnetic activity level \citep[\textit{e.g.}][]{Noyes1984}.
    Since a low metallicity star has a thinner convective envelope compared to a higher metallicity case, its convective turnover timescale is lowered and, for a given rotation rate, its Rossby number is increased \citep{AmardMatt2020}. 
    The higher Rossby number of metal-poor stars is then associated with a reduced magnetic activity, which would lead to a smaller detection rate.

    However, trends in photometric variability with stellar properties (such as mass or rotation rate) are complex and not well understood.
    \citet{Witzke2020} discussed the effect of chemical composition on the detectability of solar-like stars' rotation. They focused on solar rotation period detection for solar mass stars. They show in particular that metal-enriched stars are dominated by facular brightening hence increasing the detectability of rotation periods, while metal-poor stars show darker spots, also increasing the detectability of periods. They conclude that there should be a minimal detection probability for solar metallicity stars.

    It has been theorised by \citet{Metcalfe2016} and estimated by \citet{Morris2020} that starspots would disappear ---possibly associated to more and more frequent "Maunder-like" minimum--- as a star ages, spins down and loses its ability to sustain a convective dynamo, leaving detection of rotation periods available mostly to young stars.
    They proposed that magnetic spots would be less and less present up to disappearing and the faculae would become dominant at older ages. 
    Since high metallicity stars variability was shown to be dominated by facular brightening \citep{Witzke2020}, this might predict that they would be more easily detectable at older ages and thus at longer rotation periods.
    
    In a series of papers, \citet{Zhang2020,Zhang2020b} and \citet{Reinhold2020} showed the effect on the activity of various stellar fundamental parameters, among which the metallicity. Their sample is very similar to ours, although without necessarily detected rotation period, and more focused on near solar stars in the case of \cite{Zhang2020b} and \cite{Reinhold2020}. In particular, \cite{Reinhold2020} shows that the photometric activity slightly increases with metallicity on a narrow rotation period and temperature range around the Solar values. 
    If these trends extend to stars with observed rotation period, it would confirm claims by \citet{AmardMatt2020} that metal-enriched stars generate more intense magnetic activity than more metal-poor stars. 
    Assuming that activity decreases with increasing age ---and thus increasing rotation period--- stars with a high metallicity have a similar activity level than an older metal-poor star with a larger rotation period. 
    Since the activity level is directly correlated to the detection rate, there should be a higher detection rate of older, slow rotating stars for metal enriched population.
    
    In conclusion, it is not yet entirely clear how metallicity should affect the detection of rotation periods, but it seems likely that it will have some affect.
    At the same time, if stellar spot properties do depend on metallicity, this probably means that other stellar magnetic properties do as well.
    And if the stellar magnetic activity is affected by metallicity, it is likely that the stellar wind properties and resulting torques are also affected.
    The two may go hand in hand.
    Clearly, fully interpreting the observed distributions might be very complex (e.g., involving understanding the dependence of completeness on metallicity at the same time as correctly modelling the effect of metallicity on stellar spin-down).

\section{Conclusions}
\label{sec:conclusion}

In this work we present a sample of 4055 single main-sequence stars with Gaia DR2 Parallaxes, rotation periods from \textit{Kepler} and spectroscopic metallicities from LAMOST and APOGEE.
We displayed the sample in the colour-magnitude diagram and revealed a broad main sequence with a strong gradient of metallicity perpendicular to the main sequence. This is well-understood, and the broadening is beautifully explained by variations in metallicity (and cannot be explained by age variations; see appendix), when compared to isochrones at different metallicities.
Second, the rotation period strongly correlates with the position along the main sequence with fainter stars being much slower rotators. 
This is also well established, as shown by \textit{e.g.} \citet{Matt2015}; lower-mass stars experience a more effective magnetic torque during their main sequence, so they are spun down more efficiently and reach longer rotation periods.

Last but not least, the rotation rate weakly correlates with the position orthogonal to the main sequence. This was already seen by \citetalias{DavenportCovey2018}, who showed it could not be explained as an evolution effect alone.
Here, we established that this trend is due to a correlation between the rotation period and metallicity, with metal rich stars rotating more slowly, on average, than metal poor stars.
We suggest that the observed correlation may be influenced by metallicity-dependent detection bias (see \S~\ref{sec:detectionbias}), but it appears to be evidence for metallicity-dependent magnetic braking (\S~\ref{sec:braking}).

We compared the observed trend with predictions from spin evolution models that account for the influence of metallicity on stellar structural evolution and Rossby number, which affect the magnetic braking torques.
Detailed comparison to the observed sample is not possible, primarily due to the lack of age information, but the models predict a correlation between metallicity and rotation period that is in the same direction and of the same magnitude as what is observed.
The spin evolution models still do not include all possible effects of metallicity, and much work is needed to explore this new dimension in stellar rotation evolution.

At the same time, we anticipate much more empirical information about metallicity and rotation in the near future.
Samples include the upcoming 200 000 stars from the extended \textit{Kepler} mission, K2 \citep{Howell2014,Gordon2020} and the Gaia rotation periods extracted from DR2. 
The latter already yield about 140 000 rotation periods, but the \emph{Gaia} mission once completed is estimated to provide a sample of about 2 to 30 million stars with photometric rotation periods \citep{Lanzafame2018}. 
In the coming years, reliable asteroseismic ages as well as rotation periods of close low-mass stars may become available all over the sky with TESS \citep{TESS2015} and later PLATO \citep{PLATO2014} . In parallel, large spectroscopic surveys such as WEAVE \citep{WEAVE2012}, GALAH \citep{GALAH2015}, RAVE \citep{RAVE2017}, APOGEE, LAMOST and then 4MOST \citep{4MOST2012} will keep providing millions of spectra (and thus metallicities) for the same stars and allows to entirely trace the parameters responsible for the observed rotational distribution. 

\subsection*{Data availability}
The data underlying this article are available in CDS.

\section*{Acknowledgements}
We thank the referee for constructive comments and suggestions that improved the paper. We acknowledge funding from the European Research Council (ERC) under the European Union's Horizon 2020 research and innovation program (grant agreement No. 682393 AWESoMeStars). We thank the AWESoMeStars team members for useful discussions and J. Davenport for providing code and discussions. \\
\textit{Software:} TOPCAT \citep{Topcat}, Python, Astropy \citep{Astropy2018}, IPython \textit{IPython}, Matplotlib \citep{Matplotlib}, Numpy \citep{Numpy}, Scipy \citep{Scipy}.





\bibliographystyle{mnras}
\bibliography{CMPZD_bib.bib} 




\appendix
\section{Effects of age on the synthetic population}

The age is usually the main unknown when dealing with a stellar system, so when modelling a whole distribution, we have to make some assumptions that will inevitably strongly affect the rotation period distribution. In the corpus of the paper we chose to model a single 3 Gyr age population. In this appendix we explore the age sensitivity of our modelling of the sample by testing different age distributions.
The two main conclusions here are that an age spread (1) cannot explain the width of the main-sequence in the CMDs, leaving metallicity as the natural explanation, and (2) will serve to broaden the period distributions, while the trend of period with mass and metallicity persists.

\begin{figure*}
    \centering
    \includegraphics[width=\linewidth]{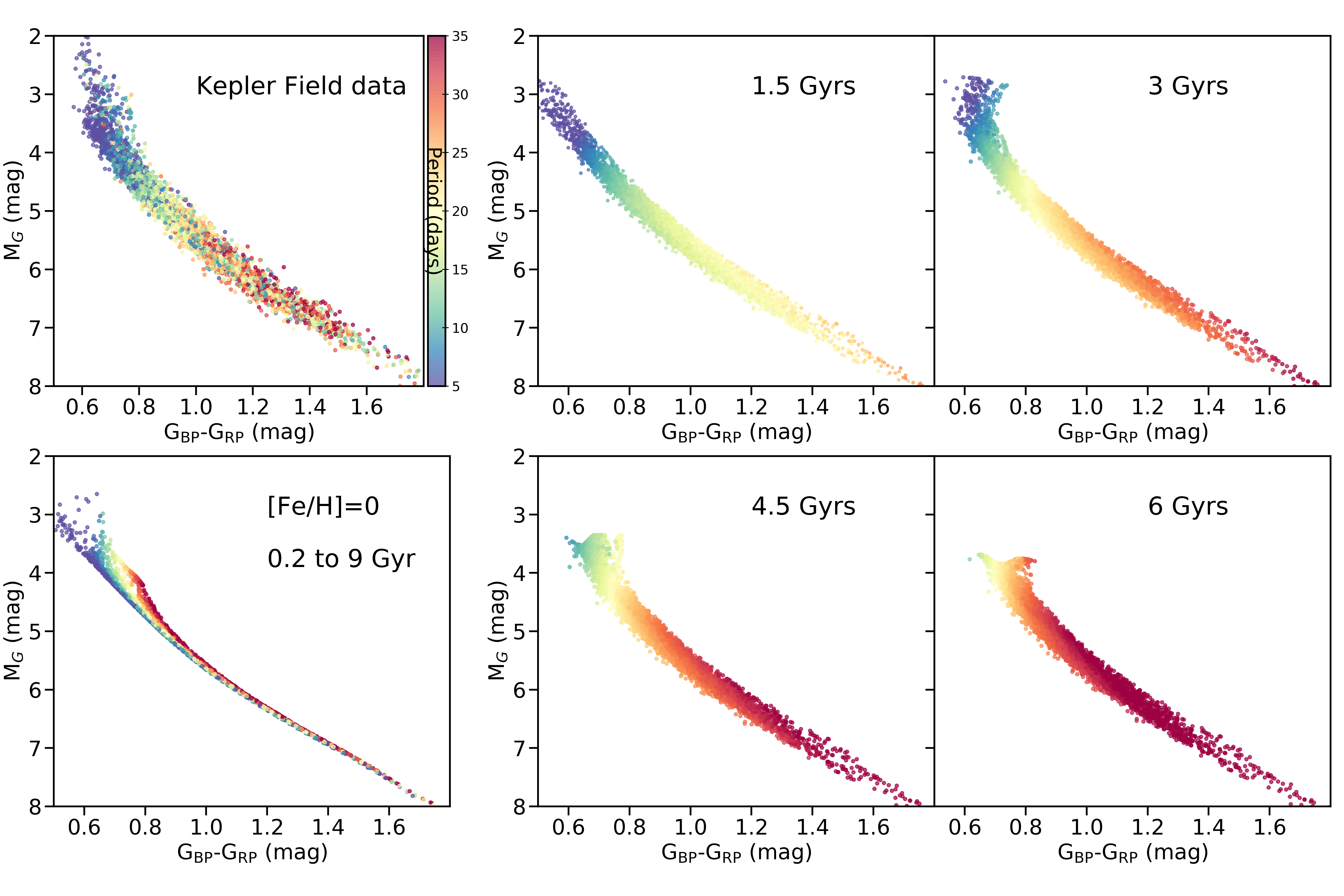}
    \caption{Colour magnitude diagram of the Kepler distribution, coloured by rotation period for the observed main sequence distribution (top left; same data as in Fig.~\ref{fig:CMD_MS}), a synthetic sample at solar metallicity but ages from 200 Myr to 9 Gyr (bottom left), and four synthetic sample similar to the one described in Sect. \ref{sec:prop_synth_sample}, with a broad range of metallicity but with ages 1.5, 3, 4.5, and 6 Gyr (respectively top-middle, top-right, bottom-middle, and bottom-right panels). The colour bar showing the rotation period is the same for all the diagrams.}
    \label{fig:multiage_CMD}
\end{figure*}

The observed rotation period distribution shown on the top left panel of the panel fig.~\ref{fig:multiage_CMD} is expected to be composed of stars with a broad range of ages with various dependence on mass and metallicity. 
On one hand, the bright blue fast rotating part of the observed CMD is only visible on the synthetic 1.5 and 3 Gyr samples, and appears much slower (redder) at older ages. 
On the other hand, the faint slow rotating red part is not visible at 1.5 Gyr because the stars have not slowed down enough, and their rotation period is still limited to about 30 days at 3 Gyr. At older ages though (4.5 and 6 Gyr), the lower end of the distribution is much more visible. 
The observed sample can be seen as a composite image of these diagrams and one could estimate a rotational age distribution of the sample by comparing different parts of the observed CMD with modelled populations.
According to \citet{VanSaders2019}, F-stars rotation period distribution is affected beyond 1.5 Gyr while G-type stars are affected only from about 4.2 Gyr. Beyond these ages, they find that the angular momentum loss of stars ceases and stellar activity becomes very minimal, leading to a different detection rate and biasing the distribution towards younger rotational ages (see also \S~\ref{sec:detectionbias}). 
A mass-metallicity-age-rotation dependant cut-off, beyond which stars are not observed or stop spinning down, could indeed explain the ages of the different parts of CMD. Thus, further work including for example a reduced braking beyond a certain Rossby threshold could be very relevant to better understand the shape of the observed \textit{Kepler} field rotation period distribution in a CMD (top left panel).

The bottom left panel of figure~\ref{fig:multiage_CMD} displays a CMD of a stellar mass distribution assuming all stars are solar metallicity with a random age between 0.2 and 9 Gyr. 
Although it does not reproduce the rotation period distribution, the age spread can partially explain the broadening of the upper main sequence. However, the shape of the lower main sequence (M$_\textrm{G} \gtrsim 5$) is almost not affected by an age spread.
It is true that age is a major unknown component, especially when it comes to rotation period evolution, but even a very large range of ages cannot explain the broadening of the main sequence or the shape of the rotation period distribution in the CMD.

\begin{figure*}
    \centering
    \includegraphics[width=\linewidth]{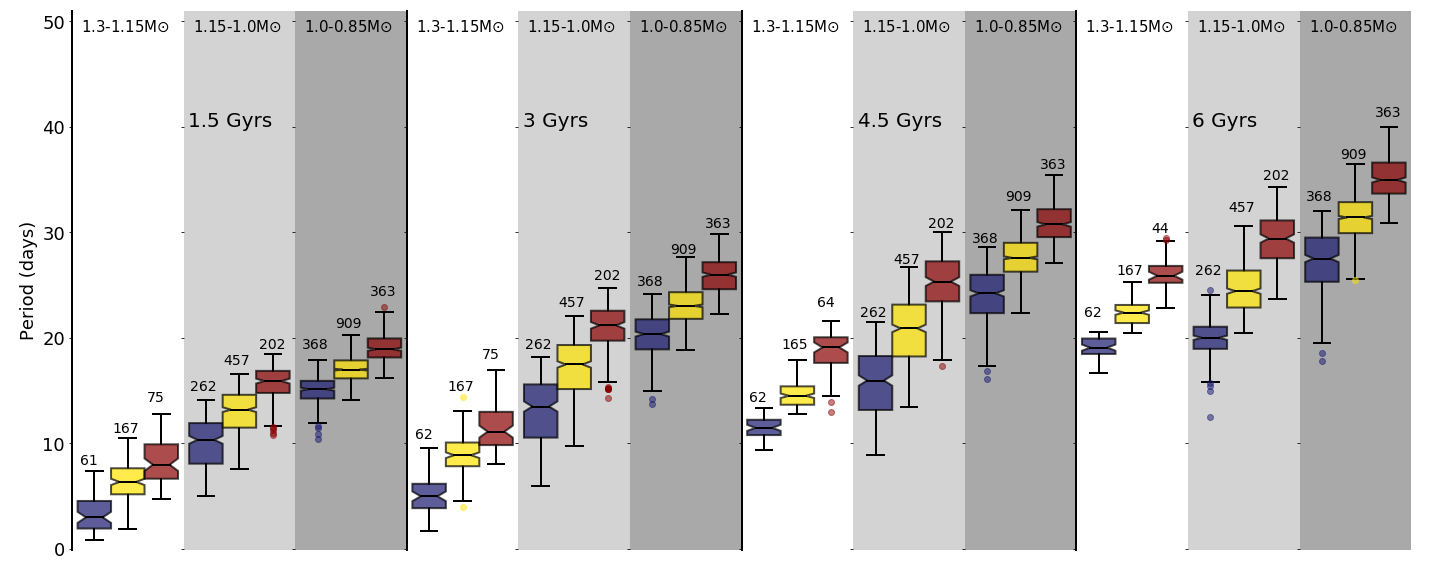}
    \caption{Same boxplots as fig~\ref{fig:Boxplot} for synthetic samples of 1.5, 3, 4.5 and 6 Gyr (from left to right).}
    \label{fig:multiage_boxplot}
\end{figure*}

Figure~\ref{fig:multiage_boxplot} shows the rotation period distribution of each mass and metallicity bin, in the same manner as Fig.~\ref{fig:Boxplot}, but at four different ages.
It is clear that, while the rotation periods in each sub-sample increase with age, the rotation period distributions display the same qualitative dependence on metallicity and mass, at all ages shown.
In each mass bin, low-metallicity stars are always spinning faster than solar-metallicity stars, while metal-rich stars are systematically slower rotators, in agreement with \citet{AmardMatt2020}. 
To compare with the observed distributions in the left panel of Figure~\ref{fig:Boxplot}, we selected the 3-Gyr sample to roughly best represent the median periods observed.
However, the synthetic period distributions are much narrower than observed, due to the singular age.
It is clear that a population of stars with a range of ages should have a broader distribution of periods, while still maintaining the trends with mass and metallicity, qualitatively similar to what is observed.

\bsp	
\label{lastpage}
\end{document}